\documentclass[namedreferences]{SolarPhysics}
\pdfoutput=1
\usepackage[optionalrh,solaenum,linksfromyear]{spr-sola-addons}	
\usepackage{graphicx}		
\usepackage{epstopdf}
\usepackage{color}		
\usepackage{url}		
 
\usepackage{amsmath}	
\usepackage{hyphenat}
\usepackage[breaklinks]{hyperref}
\usepackage[all]{hypcap}

\newcommand{\etal}{{\it et al.}}

\newcommand\arcmin{\mbox{$^{\prime}$}}
\newcommand\arcsec{\mbox{$^{\prime\prime}$}}

\begin{document}

\begin{article}

\begin{opening}

\title{Origin of the submillimeter radio emission during the time-extended phase of a solar flare}

\author{G.~\surname{Trottet}$^{1,2}$\sep
	J.-P.~\surname{Raulin}$^{2}$\sep
	G. ~\surname{Gim\'enez de Castro}$^{2}$\sep
	T.~\surname{L\"uthi}$^{3}$\sep
	A.~\surname{Caspi}$^{4\dagger}$\sep
	C.H.~\surname{Mandrini}$^{5,6}$\sep
	M.L.~\surname{Luoni}$^{5}$\sep
	P.~\surname{Kaufmann}$^{2,7}$
}
\runningauthor{G. Trottet \etal}
\runningtitle{Origin of flare time-extended submillimeter emission}

\institute{
	$^{1}$ Observatoire de Paris, LESIA-CNRS UMR 8109, Univ. P~\&~M Curie and Paris-Diderot, Observatoire de Meudon, 92195 Meudon, France;
		email:~\url{gerard.trottet@obspm.fr}\\ 
	$^{2}$ CRAAM Universidade Presbiteriana Mackenzie, S\~ao Paulo, Brazil;
		email:~\url{raulin@craam.mackenzie.br};
		email:~\url{guigue@craam.mackenzie.br};
		email:~\url{kaufmann@craam.mackenzie.br}\\
	$^{3}$ Leica Geosystem AG, Hexagon Metrology, Moenchmattweg 5, 5035 Unterenfelden, Switzerland;
		email:\url{thomas.Luethi@leica-geosystems.com}\\
	$^{4}$ Space Sciences Laboratory, University of California, Berkeley, CA, 94720-7450, USA;\\
	$^{\dagger}$ Now at: Laboratory for Atmospheric and Space Physics, University of Colorado, Boulder, CO, 80303, USA;
		email:~\url{amir.caspi@lasp.colorado.edu}\\
	$^{5}$ Instituto de Astronom\'\i a y F\'\i sica del Espacio, CONICET-UBA, CC. 67 Suc. 28, 1428, Buenos Aires, Argentina;
		email:~\url{mandrini@iafe.uba.ar};
		email:~\url{mluoni@iafe.uba.ar}\\
	$^{6}$ Facultad de Ciencias Exactas y Naturales, FCEN-UBA, Buenos Aires, Argentina\\
	$^{7}$ Centro de Componentes Semicondutores, Universidade Estadual de Campinas, 
	Campinas, Brasil\\
}

\begin{abstract}
Solar flares observed in the 200--400~GHz radio domain may exhibit a slowly varying and time-extended component which follows a short (few minutes) impulsive phase and which lasts for a few tens of minutes to more than one hour. The few examples discussed in the literature indicate that such long-lasting submillimeter emission is most likely thermal bremsstrahlung. We  present  a detailed analysis of the time-extended phase of the 2003~October~27 (M6.7) flare, combining  1--345~GHz total-flux radio measurements with X-ray, EUV, and H$\alpha$ observations.  We find that the time-extended radio emission is, as expected, radiated by thermal bremsstrahlung.  Up to 230~GHz,  it  is entirely produced in the corona by hot and cool materials at  7--16~MK and 1--3~MK, respectively. At 345~GHz, there is an additional contribution from chromospheric material at a few 10$^4$~K. These results, which may also apply to other millimeter--submillimeter radio events, are not consistent with the expectations from standard semi-empirical models of the chromosphere and transition region during flares, which predict observable radio emission from the chromosphere at all frequencies where the corona is transparent.
\end{abstract}
\end{opening}


\section{Introduction}
Since 2000, new instrumentation has allowed us to observe solar flares at submillimeter wavelengths. Such observations are routinely carried out by the {\it Solar Submillimeter Telescope} (SST; \opencite{Kau:al-08}) at 212 and 405~GHz, and studies of ten major bursts detected by this instrument have been reported in the literature. Measurements at 230 and 345~GHz have been obtained for two more flares by the {\it K\"oln Observatory for Submillimeter and Millimeter Astronomy} (KOSMA; \opencite{Lut:al-04}; \opencite{Luth:al-04}) telescope, one of them being also observed at 210~GHz by the {\it Bernese Multibeam Radiometer for KOSMA} (BEMRAK).

Above 200~GHz, these events exhibit an impulsive phase lasting for a few minutes, which is sometimes followed by slowly varying and time-extended emission (the ``gradual'' phase). While, for some events, the $>$200~GHz emission appears as the extension toward high frequencies of the gyrosynchrotron emission seen in the microwave domain \cite{Tro:al-02,Lut:al-04,Rau:al-04,Gim:al-09}, other events exhibit an unexpected upturn towards the THz domain (e.g., \opencite{Kau:al-04}). Such spectra with positive slopes in the millimeter--submillimeter domain have been measured during both the impulsive phase \cite{Kau:al-04,Sil:al-07,Kau:al-09} and the gradual phase \cite{Tro:al-02,Lut:al-04,Luth:al-04}.

Although various theoretical ideas have appeared in the literature (e.g., \opencite{Kau:Rau-06}; \opencite{Fle:Kon-10}), the emission mechanism responsible for a positive slope in the sub-THz domain remains uncertain, at least during the impulsive phase. During the gradual phase, it has been shown that thermal bremsstrahlung from the chromosphere and the corona may account for the observed radio spectrum, particularly during the late decay of the flare. This is the case for the 2000~March~22 (X1.1) flare \cite{Tro:al-02} and for the 2003~October~28 (X17.2) flare \cite{Tro:al-08}, where the radio spectrum, late in the gradual phase, is roughly flat below 200~GHz and increases at higher frequencies. Such spectra were interpreted as optically-thin thermal bremsstrahlung from the corona below $\sim$200~GHz, and it was suggested that the excess emission measured at higher frequencies arose from thermal bremsstrahlung from the lower atmosphere.

The best example of a thermal phase at 230 and 345~GHz reported so far is the time-extended component of the 2001 April 12 (X2.1) flare, which lasts for more than two hours and during which the radio spectrum is flat first up to 230~GHz and later up to 89~GHz, with a positive slope above these frequencies \cite{Lut:al-04}. However, although the radio emission during the gradual phase has been conjectured to arise from separate contributions of coronal and chromospheric sources, the few available observations  have not allowed a  characterization of the plasmas involved.
 
This paper presents an analysis of radio observations of the 2003~October~27 (M6.7) flare at $\sim$12:30~UT obtained in the 1--405~GHz range by the USAF Radio Solar Telescope Network\footnote{\url{http://www.ngdc.noaa.gov/stp/solar/solarradio.html}} (RSTN), the Bumishus patrol telescopes (Institute of Applied Physics of Bern University), BEMRAK, the KOSMA telescope, and the SST. Above 200~GHz, the radio event is similar to the 2001 April 12 radio burst. Indeed: (i) it exhibits a short impulsive phase followed by a time-extended component; (ii) the impulsive radio emission is gyrosynchrotron radiation from relativistic electrons; and (iii) the gradual phase is optically-thin (8--230~GHz) and optically-thick (345~GHz) thermal bremsstrahlung. The main goal of this study is to combine the radio observations with soft X-ray, UV/EUV and optical data in order to discuss the origins of the thermal radio emission during the gradual phase. The paper is organized as follows. Section~\ref{Sec_Obs} describes the radio observations and flux calibration of the SST, KOSMA and BEMRAK measurements. The results are discussed in Section~\ref{Sec_Results}, and conclusions are drawn in the final Section.


\section{Observations and Data Analysis}
\label{Sec_Obs}

\begin{figure}
\centerline {
	\includegraphics[width=0.9 \textwidth]{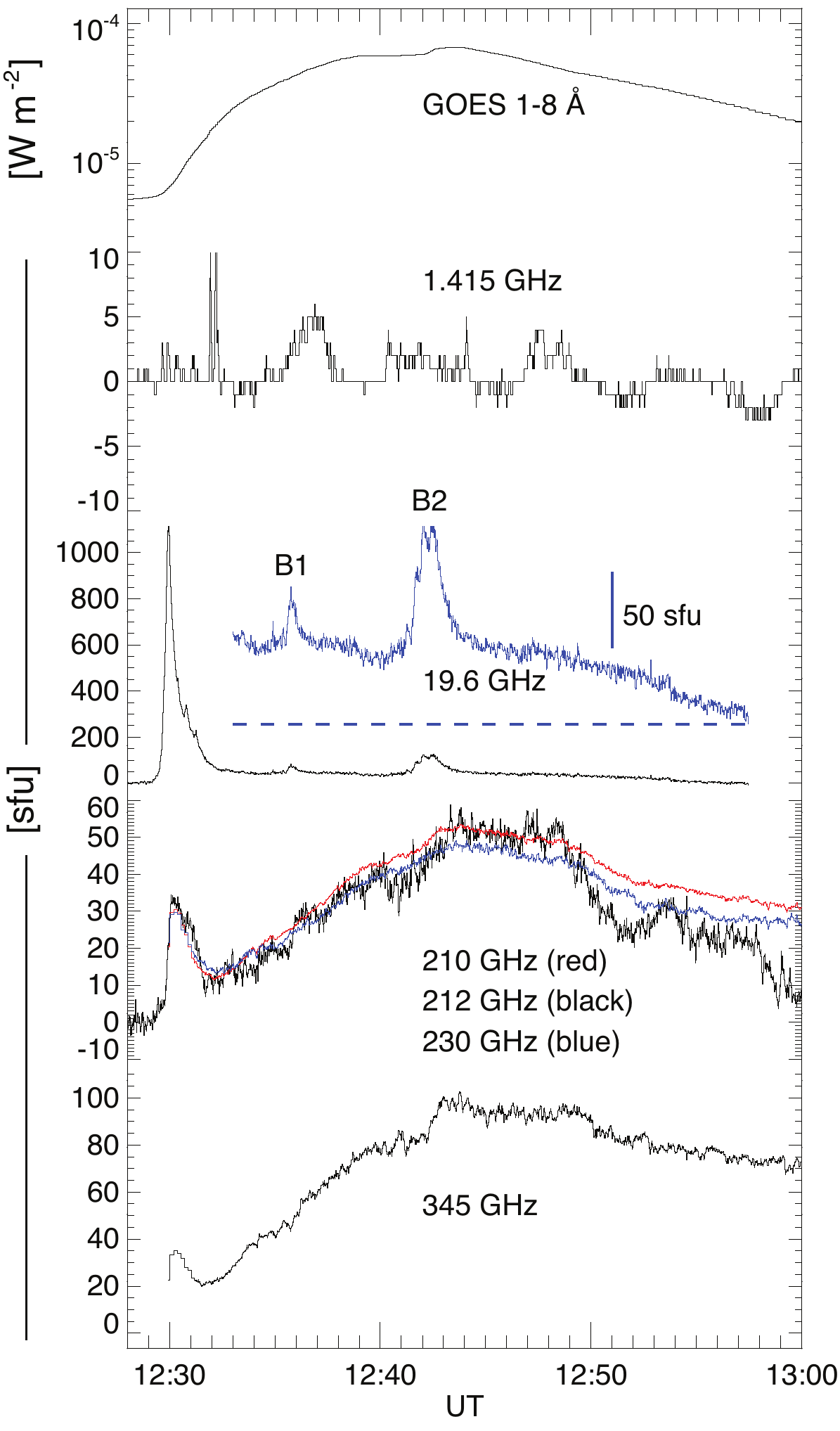}
}
\caption[]{
Time histories of X-ray and radio emissions during the 2003~October~27 flare at $\sim$12:30~UT. From top to bottom: 1-8~{\AA} soft X-ray flux (background subtracted) from the GOES-12 X-ray Sensor; radio flux densities at 1.415~GHz from RSTN; at 19.6~GHz from the Bumishus telescopes (the blue curve shows the time extended phase with an enlarged flux density scale); at 210~GHz from BEMRAK (red), 230~GHz from KOSMA (blue), and at 212~GHz from SST (black); and at 345~GHz from KOSMA.
}
\label{fig_lcradio}
\end{figure}

The 2003~October~27 radio event at $\sim$12:30~UT was associated with a GOES M6.7-class soft X-ray (SXR) flare and an H$\alpha$ sub-flare which occured in active region AR~10486 at S17~E25. The flare was well-observed in the UV and EUV, from $\sim$12:28 to $\sim$13:00~UT, by the {\it Transition Region and Coronal Explorer}\footnote{\url{http://trace.lmsal.com/}} (TRACE; \opencite{Han:al-99}). Unfortunately, only the late part of the flare (after $\sim$12:46~UT) was observed in the hard X-ray domain ($>$20 keV), by the {\it Reuven Ramaty High Energy Solar Spectroscopic Imager}\footnote{\url{http://hesperia.gsfc.nasa.gov/hessi}} (RHESSI; \opencite{Lin:al-02}). Radio observations were available over a wide spectral domain from 10~kHz up to 405~GHz. Figure~\ref{fig_lcradio} shows the time evolution of the GOES 1--8~{\AA} SXR flux and of the radio flux densities\footnote{1 sfu = 10$^{-22}~$W~m$^{-2}$~Hz$^{-1}$} at selected frequencies in the 1--400~GHz domain. Total flux densities have been obtained by RSTN at eight frequencies in the 0.245--15~GHz range and by the Bumishus patrol telescopes at 8.4, 11.8, 19.6, 35, and 50~GHz (until 12:57~UT). Measurements at 212 and 405~GHz were made by the SST. KOSMA and BEMRAK provided measurements at 230 and 345~GHz, and at 210~GHz, respectively, until 13:00~UT.

The radio event exhibits two phases: an impulsive burst which lasts for about 3~min ($\sim$12:29--12:32~UT) and a slowly varying and time-extended emission which lasts up to about 13:15~UT at 212~GHz. It should be noted that: (i)~although this time-extended emission is more pronounced and larger than the impulsive burst in the 200--400~GHz range, it is also observed at all other frequencies above 1.415~GHz, except at 50~GHz due to strong atmospheric absorption; and (ii)~at 212~GHz, the slowly-varying emission starts slightly before the impulsive rise, simultaneously with the SXR emission. Such a time sequence is similar to that observed in other events with $>$200~GHz emission \cite{Lut:al-04,Tro:al-08,Gim:al-09}. At microwave frequencies, two impulsive bursts, B1 and B2, with maxima at $\sim$12:35:49~UT and $\sim$12:42:20~UT, respectively, are superimposed on the slowly-varying component.

Figure~\ref{fig_lcradio} shows that no significant emission is measured at 1.415~GHz. Although radio activity was observed in the decimeter--hectometer range (see Radio Monitoring\footnote{\url{http://secchirh.obspm.fr/survey.php?dayofyear=20031027&composite=1} and\\ \url{http://secchirh.obspm.fr/survey.php?dayofyear=20031027&composite=2}}), these radio emissions are not associated with AR~10486, so the flare under study is radio-silent at frequencies below $\sim$2~GHz. Indeed:

\begin{itemize}
	\item [--] There is a permanent noise storm which exhibits an enhancement around 12:30~UT, i.e., close to the maximum of the impulsive phase of the event under study. Radio imaging by the {\it Nan\c{c}ay Radioheliograph} (NRH) in the 150--450~MHz range shows that this noise storm activity is spatially associated with AR~10488 (at N09~E12), not with AR~10486. A decameter--hectometer Type~III storm is also associated with this noise storm, as is sometimes the case (see \opencite{Elg-77}).
	\\
	\item [--] \begin{sloppypar}Between 12:30 and 14:00~UT, there are some decimeter--hectometer Type~III bursts, indicating that some electron beams accelerated in the low corona reach interplanetary space. Here, again, NRH images show that these Type~IIIs are not associated with the flare under study, but rather with activity from AR~10484 (at N07~W46).\end{sloppypar}
\end{itemize}

The 2003~October~27 flare thus belongs to the class of ``confined microwave events'' \cite{Kle:al-11}, where microwave-emitting electrons remain confined in the low corona. This is consistent with the analysis of the magnetic field topology of this flare by \inlinecite{Luo:al-07}, which indicates that energy release occurred in compact loops at a magnetic null very low in the corona ($\sim$3~Mm above the photosphere). 


\subsection{Flux Densities at 212 and 405 GHz}
\label{Sec_FluxSST}

\begin{figure}
\centerline {
	\includegraphics[width=0.9 \textwidth]{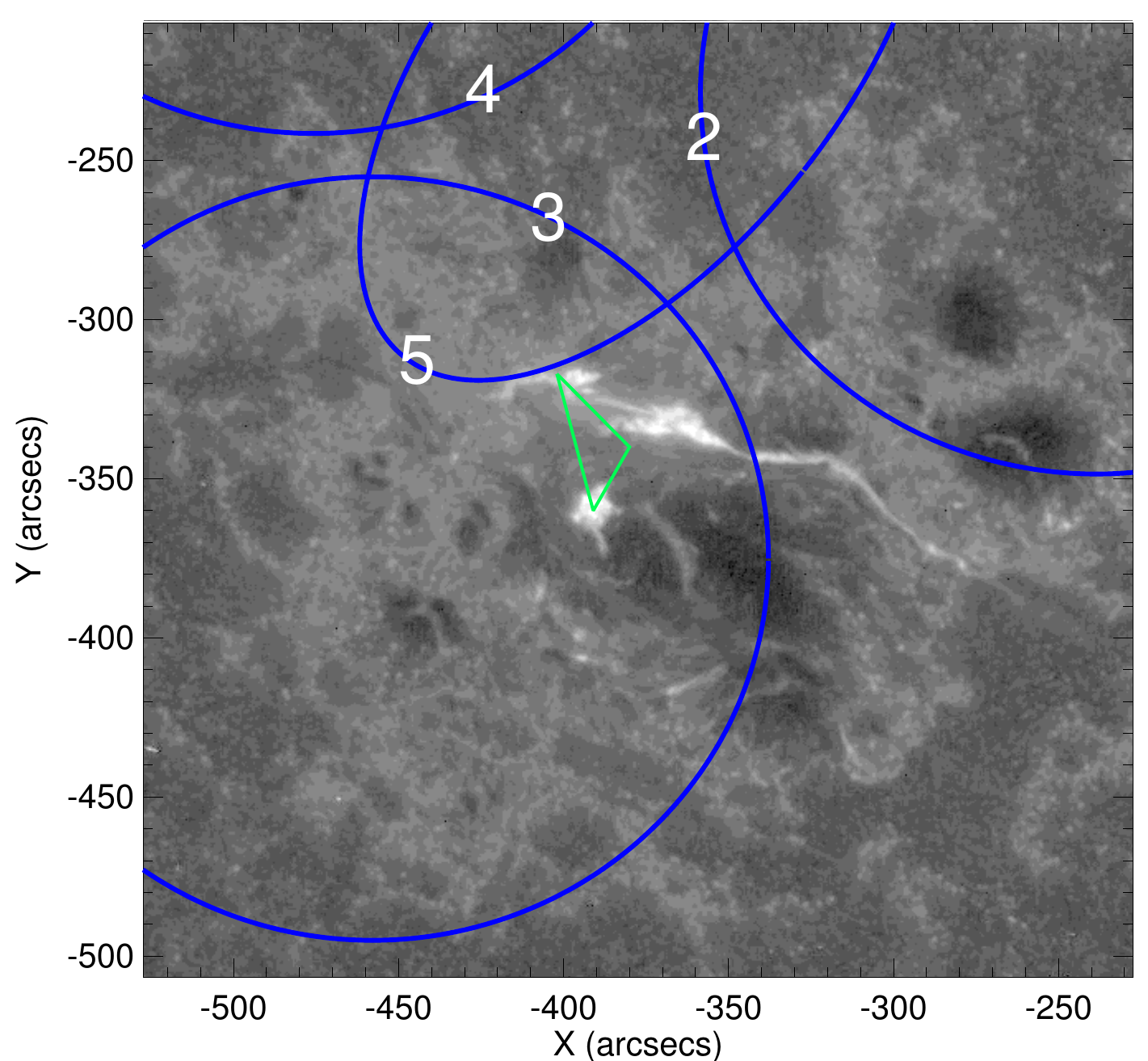}
}
\caption[]{
Configuration of the SST beams during the flare. The circles and the ellipse represent the 50\% contour (HPBW) of beams 2--5, overlaid on a TRACE 1600~{\AA} image taken at 12:46:21~UT. The numbers indicate the beam. The green triangle represents the region where the 212 and 405~GHz flare sources may be located without being detected by beam~2 at 212~GHz or by beam~5 at 405~GHz (see text).}
\label{fig_BeamPos}
\end{figure}

The SST observed the 2003~October~27 event with five independent total-power receivers: channels 1--4 at 212~GHz and channel 5 at 405~GHz. Figure~\ref{fig_BeamPos} shows the 50\%-level contours for beams~2--5 while SST was tracking AR~10486, overlaid on a TRACE 1600~{\AA} image obtained at $\sim$12:46~UT, close to the maximum of the time-extended phase. Beam~1 is outside the field of view of the Figure. At 212~GHz, beams~1--4 may be approximated by circular Gaussians with a half-power beam-width (HPBW) of $\sim$4\arcmin. The 405~GHz channel was, for this period, oblate, and is better represented by an elliptical Gaussian with a HPBW of $\sim$$2\arcmin \times 4\arcmin$.

\begin{figure}
\centerline {
	\includegraphics[width=0.9 \textwidth]{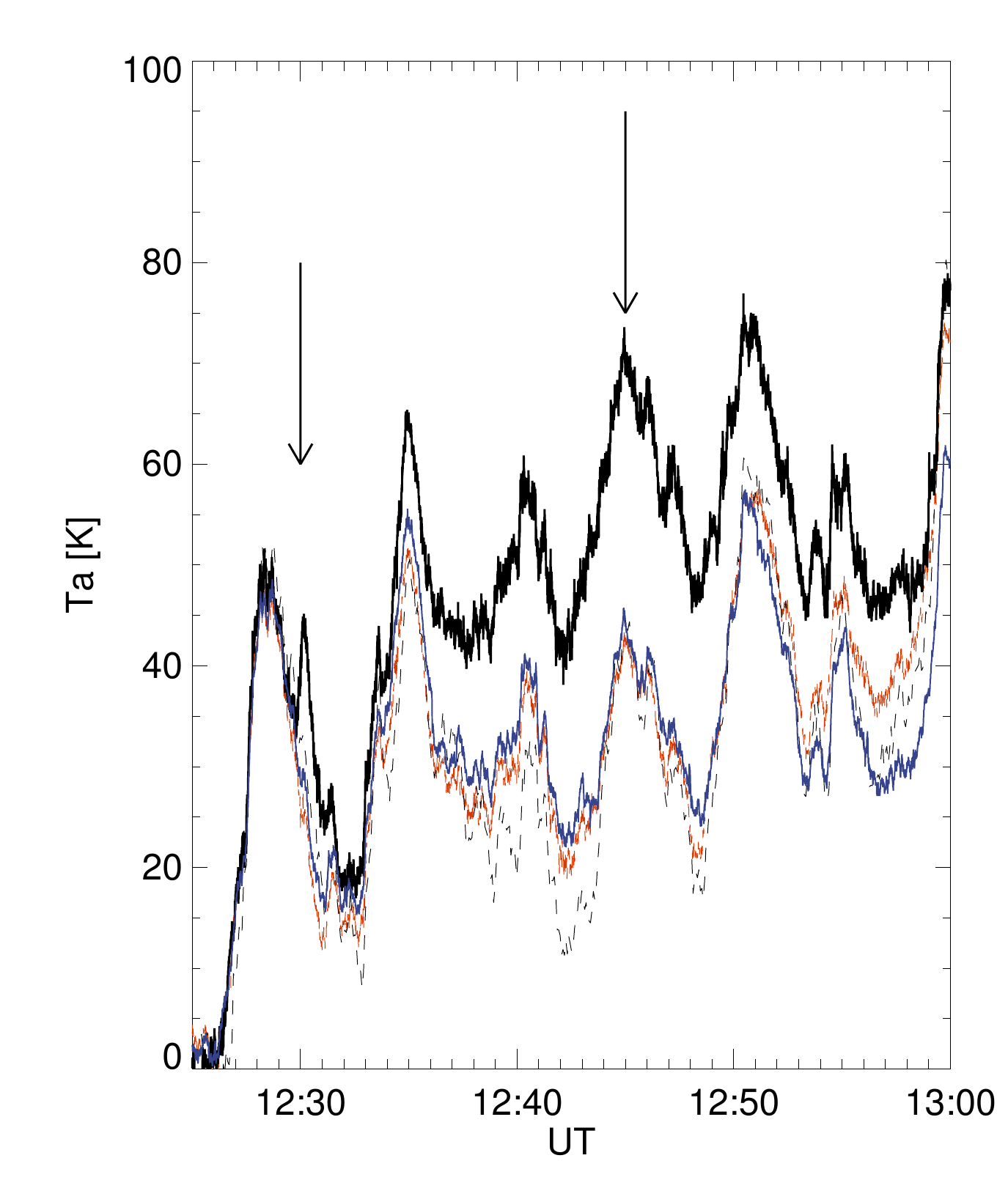}
}
\caption[]{ 
Antenna temperatures observed by the different SST beams at 212~GHz (beam~2: red; beam~3: thick black; beam~4: dashed black) and at 405~GHz (beam~5: blue). Arrows indicate the maxima of the flare impulsive and extended phases at 12:30~UT and 12:45~UT, respectively.}
\label{fig_sstbeams}
\end{figure}

Due to the low solar elevation angle $\theta_{\rm el} \approx 35^\circ$ at the time of the flare, along with the rather large zenith opacity, the antenna temperature time profiles at 212 and 405~GHz are quite similar, dominated by atmospheric fluctuations as illustrated in Figure~\ref{fig_sstbeams}. However, we note that the signal from beam~3 shows additional temporal features compared to those in beams~2, 4 and 5. In particular, we observe a rapid variation at 12:30~UT followed by a gradual increase between 12:32 and 13:00~UT, which correspond, respectively, to the impulsive and extended phases of the flare, shown in Figure~\ref{fig_lcradio}. Therefore, the flare was detected mainly within beam~3, with negligible contributions in beams~2, 4 and 5. This is consistent with the beam positions relative to the UV flaring structures, as shown in Figure~\ref{fig_BeamPos}.

Since the signal from beam~4 is very noisy, including spiky interferences, we instead consider that the signal from beam~2 is the one which is characteristic of the emission of the whole active region. Figure~\ref{fig_sstbeams} shows that, before the flare, the signals from beams~2 and 3 are identical.  This indicates that the pre-flare (background) signal does not depend critically on the position of the whole active region within the beam. The signal of beam~2 has thus been subtracted from that of beam~3 in order to suppress variations due to atmospheric fluctuations  and to obtain the excess antenna temperature $T_a$ due to the flare.

The flux density at  212~GHz, $S_{212}$, is then given by:
\begin{equation}
	S_{212} = \frac{2k_{\rm B} T_a}{A_a\eta_a} \exp{\left(\tau_{\rm z}/\sin{(\theta_{\rm el})}\right)} \,,
\label{eq:calibration}
\end{equation}
where $k_{\rm B}$ is the Boltzmann constant, $A_a$ is the antenna surface area, $\eta_a$ the aperture efficiency, and $\tau_{\rm z}$ the zenith 
opacity of the Earth's atmosphere.  We estimated $\tau_{\rm z(212)}\approx0.46$ and $\tau_{\rm z(405)}\approx1.9$ at 212 and 405~GHz, respectively, 
using the method described in \inlinecite{Mel:al-05}. For the present observation, the antenna aperture efficiency $\eta_a$ is 20$\%$ 
at 212~GHz. The main sources of uncertainty are temporal fluctuations of $\tau_{\rm z}$ and the aperture efficiency, which result in 
a total uncertainty of 20\%. Figure~\ref{fig_lcradio} shows the 212~GHz flux density as computed using Equation~(\ref{eq:calibration}). 

In Equation~(\ref{eq:calibration}), we have assumed that the source is located over the center of beam~3. In order to account for the actual source location, we need to multiply $S_{212}$ by the factor $\exp(\theta_{\mathrm{off}}^2/2\sigma^2)$. Here, $\theta_{\mathrm{off}}$ is the angular distance between the center positions of the source and of beam~3, and $\sigma$ represents the standard deviation of the convolution of a circular Gaussian source by the beam. We estimated that the maximum flux density at 405~GHz should have been $\sim$125~sfu by extrapolating the increasing 212--345~GHz flux spectrum up to 405~GHz (see Figure~\ref{fig_lcradio}). However,  the event was detected only in beam~3 at 212~GHz, and was not detected in beam~5 at 405~GHz within the r.m.s. noise of 30--40~sfu.  We use these two facts  to  constrain the source size and location, and to estimate $\theta_{\mathrm{off}}$. To do so, we further supposed that the center of the 405~GHz-emitting source should be near the maximum of the SXR emission observed by RHESSI (see Section~\ref{Sec_RHESSI}) and between the UV ribbons observed by TRACE (see Figure~\ref{fig_BeamPos}). We then computed the emission that should be detected at 405~GHz by convolving a circular Gaussian source of 125~sfu with beam~5. We found that sources with sizes ranging from $10\arcsec$ to $70\arcsec$, with centers located within the green triangle in Figure~\ref{fig_BeamPos}, fulfill the above two requirements, i.e., no burst detection in beam~5 at 405~GHz or in beam~2 at 212~GHz. The correction factor for $S_{212}$ in Equation~(\ref{eq:calibration}) is thus between 25\% and 40\%, including a $20\arcsec$ indetermination in antenna pointing. At 212~GHz, the maximum uncertainties on the flux density are thus taken as -30\% and +40\%.

 
\subsection{Flux Densities at 210, 230 and 345 GHz}
\label{Sec_FluxKOSMA}

On 2003~October~27, the KOSMA target was AR~10486, where the flare under study was located. KOSMA uses a tiltable subreflector which periodically deflects the antenna beams by $6\arcmin$ in  elevation, from the flaring active region to a quiet reference region on the Sun. The ``1~second on / 0.1~second off'' observation cycle allows for compensation of atmospheric attenuation changes with time scales greater than 1~s \cite{Lut:al-04}. The absolute flux density calibration was obtained from hourly observations of the sky. Unfortunately, before 12:31~UT the subreflector was wobbling the wrong way, staying 10~s
 on the quiet reference region and 0.1~s on AR~10486. With such a short sampling of the active region,  we could determine neither the excess flux of the pre-flare active region with respect to the quiet reference flux, nor the onset of the flare itself. Significant flux  density excesses at 210, 230 and 345~GHz relative to the quiet reference region were thus measured only after 12:29:56~UT, i.e., slightly before the maximum of the impulsive phase at 212~GHz (see Figure~\ref{fig_lcradio}). 

In order to estimate the net flux density excess due to the flare alone, we proceeded as follows. We first assumed that the flare emission started at the same time at 210, 230 and 345~GHz as at 212~GHz. This is generally the case for events detected in the 200--400~GHz range (e.g., \opencite{Kau:al-04}; \opencite{Lut:al-04}; \opencite{Rau:al-04}; \opencite{Kau:al-09}). The growth of the time-extended emission at 212~GHz is well-fit by a straight line between 12:32 and 12:35~UT. By extrapolating this straight line back to the time $t_0$ where it crosses the pre-event flux density level, we get an estimate of the starting time of the burst: $t_0$~=~12:38:37~UT~$\pm$~5~s. At 210 and 230~GHz, straight lines have been fit to the same time interval as at 212~GHz and extrapolated back to $t_0$, providing the actual pre-flare flux densities at 210 and 230~GHz. These are found to be $36 \pm 6$ and $47 \pm 7$~sfu at 210 and 230~GHz, respectively. These uncertainties are estimated by taking into acount the uncertainty on $t_0$, the uncertainties on the fit parameters, and the amplitude of the fluctuations of the measured flux density. Figure~\ref{fig_lcradio} shows that the derived flare excesses at 210 and 230~GHz agree well with that measured at 212~GHz, justifying the procedure described above. 

At 345~GHz, the early increase of the time-extended emission cannot be fit by a single straight line. We obtained upper and lower limits for the contribution of the active region before the flare by fitting straight lines to the 345~GHz flux density excess measured between 12:32 and 12:32:34~UT, and between 12:32:34 and 12:33:50~UT, respectively, and extrapolating back to $t_0$. The contribution of the active region is thus found to be between $74 \pm 8$ and $98 \pm 10$~sfu. For the following Sections, we adopt the mean value, $86 \pm 14$~sfu. 

Following \inlinecite{Lut:al-04}, we adopt maximum uncertainties of -30\% and +40\%  at 210 and 230~GHz. An absolute error of $\pm$14 sfu is added at 345~GHz (see previous paragraph). Figure~\ref{fig_lcradio} shows that, during the time-extended phase, the time evolution of the 210--230~GHz flux and of the 345~GHz fluxes are different, with the flux density at 345~GHz being systematically larger (about 2$\times$ after 12:40~UT) than that at 210--230~GHz, even when using the upper limit of the pre-flare active region contribution. As the above uncertainties constitute conservative upper limits of the flux accuracy, we consider that the difference between the 210--230~GHz and 345~GHz flux densities is significant despite being only slightly larger than the uncertainties.


\subsection{RHESSI spectral and imaging observations}
\label{Sec_RHESSI}

We used RHESSI X-ray spectra and images to quantitatively characterize the hot ($\gsim$10 MK) thermal plasma in the 2003~October~27 flare.  While the GOES SXR data provide some information about the flare plasma temperature and emission measure (see Section \ref{Sb1_Gradual}), the broadband response over only two channels (1--8 and 0.5--4~{\AA}, corresponding to $\sim$1.6--12 and $\sim$3.1--25~keV, respectively) thus yields only a single, average measurement with no spatial information.  On the other hand, in the X-ray range ($\sim$3--100~keV), RHESSI's $\sim$1~keV FWHM spectral resolution enables precise measurements of the hottest plasma, which can then be combined with the GOES observations to also measure cooler components and to thus more accurately determine the  temperature distribution in the flare. RHESSI also provides spatial information for the hot plasma, with an angular resolution as good as $\sim$$2\arcsec$.

RHESSI observations of the 2003~October~27 flare are available only in the late decay, from $\sim$12:46 to $\sim$13:04~UT. Because of the limited observing period and the rapidly-decaying hard X-ray flux (the incident $\sim$6--25 keV flux decreases by about two orders of magnitude over these $\sim$18~minutes), we analyzed only a single time interval, from 12:47 to 12:49~UT.  The 3--100~keV spatially-integrated spectrum was accumulated over this 2-minute period using the single best detector, G4, with 1/3-keV energy bins (the instrument channel width; \opencite{Smi:al-02}), and the non-solar background was then subtracted (cf. \opencite{Cas-10}); above $\sim$3~keV, the solar pre-flare background is negligible compared to the flare emission. The spectrum decreases steeply with energy ({\it e}-folding of $\sim$2~keV), with no appreciable flare emission above $\sim$33~keV in the analyzed period.

For precise spectroscopy, we employed the low-energy instrument response calibration improvements described by \inlinecite{Cas-10}. Following the method of \inlinecite{Cas:Lin-10}, the $\sim$4.67--33~keV spectrum was forward-fit with a photon model consisting of two isothermal continua (a single isothermal did not yield an acceptable fit) and two Gaussian features (representing unresolved excitation lines of highly-ionized Fe and Ni).  The spectrum was well-fit by this model (reduced $\chi^{2}\approx0.71$) and yielded, for each of the two thermal components, a best-fit temperature and emission measure of $T_{1}\approx24$~MK, $E\!M_{1}\approx0.24\times10^{49}$~cm$^{-3}$, $T_{2}\approx16$~MK, and $E\!M_{2}\approx1.7\times10^{49}$~cm$^{-3}$, respectively.

An image at 6--30~keV was accumulated for the same time interval, using the CLEAN image reconstruction algorithm with uniform weighting \cite{Hur:al-02} and grids 3, 4, 5, 6, 8, and 9 (grid 7 was excluded due to its $\sim$20~keV low-energy threshold) to achieve a $\sim$10$\arcsec$ spatial resolution (see Section \ref{Sb1_Gradual}).  Although it is not possible to image the two thermal components separately, we can nevertheless derive their centroid positions relative to the combined emission (cf. \opencite{Cas:Lin-10}). From the spatially-integrated spectral model, the $\sim$24~MK source was found to contribute $\sim$45\%, $\sim$63\%, and $\sim$90\% of the total count flux at 6.3--7.3, 9--12, and 17--18~keV, respectively, and imaging shows that the centroid position of the emission in each energy band is correlated with this fractional contribution. From this, we extrapolated the centroid positions of the $\sim$24 and $\sim$16~MK thermal plasmas and found them to be separated by $\sim$$11\arcsec \pm 3.2\arcsec$.  The 6--30~keV source is dominated ($\gtrsim$80\%) by counts from the $\sim$24~MK component, and encompasses a projected area (cf. \opencite{Cas-10}) of $\sim$280--310 square arcsec within the 50\% contour, and $\sim$560--660 square arcsec within the 30\% contour (both values have been corrected for broadening by the instrument point-spread function).  Although the $\sim$16~MK component contributes only weakly to the imaged emission, the morphology and centroid separation suggests that it is of a similar size.


\section{Results and Discussion}
\label{Sec_Results}

\subsection{Impulsive Phase}
\label{Sec_Impulsive}
 
\begin{figure}
\centerline {
	\includegraphics[width=0.9 \textwidth]{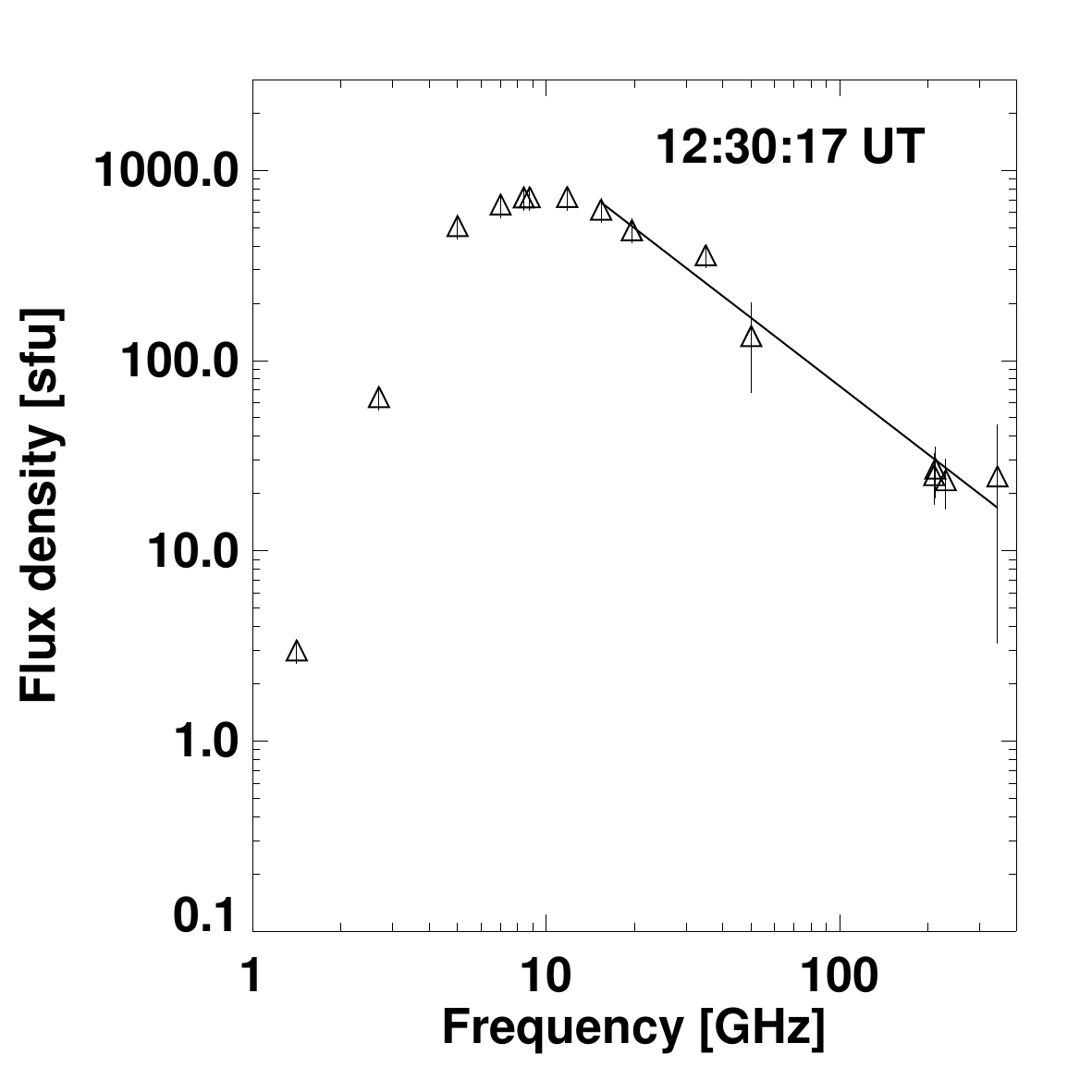}
}
\caption[]{
The radio spectrum near the maximum of the 212~GHz emission during the impulsive phase of the 2003~October~27 event. The solid line is a power-law fit to the decreasing part of the radio spectrum, with spectral index $\alpha = -1.2$.}
\label{fig_spimp}
\end{figure}

Figure~\ref{fig_spimp} displays the radio spectrum observed in the 1--400~GHz range around the maximum of the impulsive phase at 212~GHz.  This spectrum  is reminiscent of the usual gyrosynchrotron emission from  non-thermal electrons, commonly observed in the microwave domain. It peaks around 10~GHz, has a sharp low frequency cutoff between 1.415 and 2.695~GHz, and has a positive spectral index of $\sim$4 below 5~GHz. This is steeper than the index of 2.5--2.9 predicted by gyrosynchrotron self-absorption. Such steep spectra are generally ascribed to the suppression of microwave emission by the ambient medium (the Razin effect). In magnetic fields of a few hundred Gauss, the Razin effect becomes significant at a few GHz for ambient densities of at least a few 10$^{10}$~cm$^{-3}$. For stronger magnetic fields, higher ambient densities are required for the same level of suppression. Such high densities inferred for the microwave-emitting region further support the statement that the 2003~October~27 flare is a ``confined microwave event'' (see Section~\ref{Sec_Obs}).
 
The optically-thin, decreasing part of the spectrum above 15~GHz is well-represented by a power law, as shown by the solid line in Figure~\ref{fig_spimp}. The power-law spectral index $\alpha$ is about $-1.2$ around the maximum of the impulsive phase at 212~GHz. This optically-thin radiation is primarily emitted by MeV electrons. Considering the ultra-relativistic case as a gross approximation, the spectral index $\delta$ of the instantaneous distribution of radio-emitting electrons is $\delta = 2\alpha-1$ \cite{Gin:Syr-65}\footnote{\inlinecite{Gin:Syr-65} define $\alpha,\delta > 0$; we have modified their equation to eliminate this assumption, taking $\alpha \rightarrow -\alpha$ and $\delta \rightarrow -\delta$.}, thus $\delta \approx -3.4$ at the maximum of the impulsive phase. Such a hard electron spectrum is characteristic of gyrosynchrotron events observed up to millimeter--submillimeter wavelengths (e.g., \opencite{Tro:al-02}; \opencite{Lut:al-04}; \opencite{Rau:al-04}; \opencite{Gim:al-09} and references therein). 

Figure~\ref{fig_spimp} indicates that, within the large uncertainties, the emission at 210--230 and 345~GHz appears to be the high-frequency part of the gyrosynchrotron spectrum observed in the microwave domain. Such a continuation of the microwave spectrum to the 300--400~GHz domain has been previously observed, although for larger, GOES X-class flares.
 

\subsection{Gradual Phase}
\label{Sec_Gradual}

\begin{figure}
\centerline {
	\includegraphics[width=0.9 \textwidth]{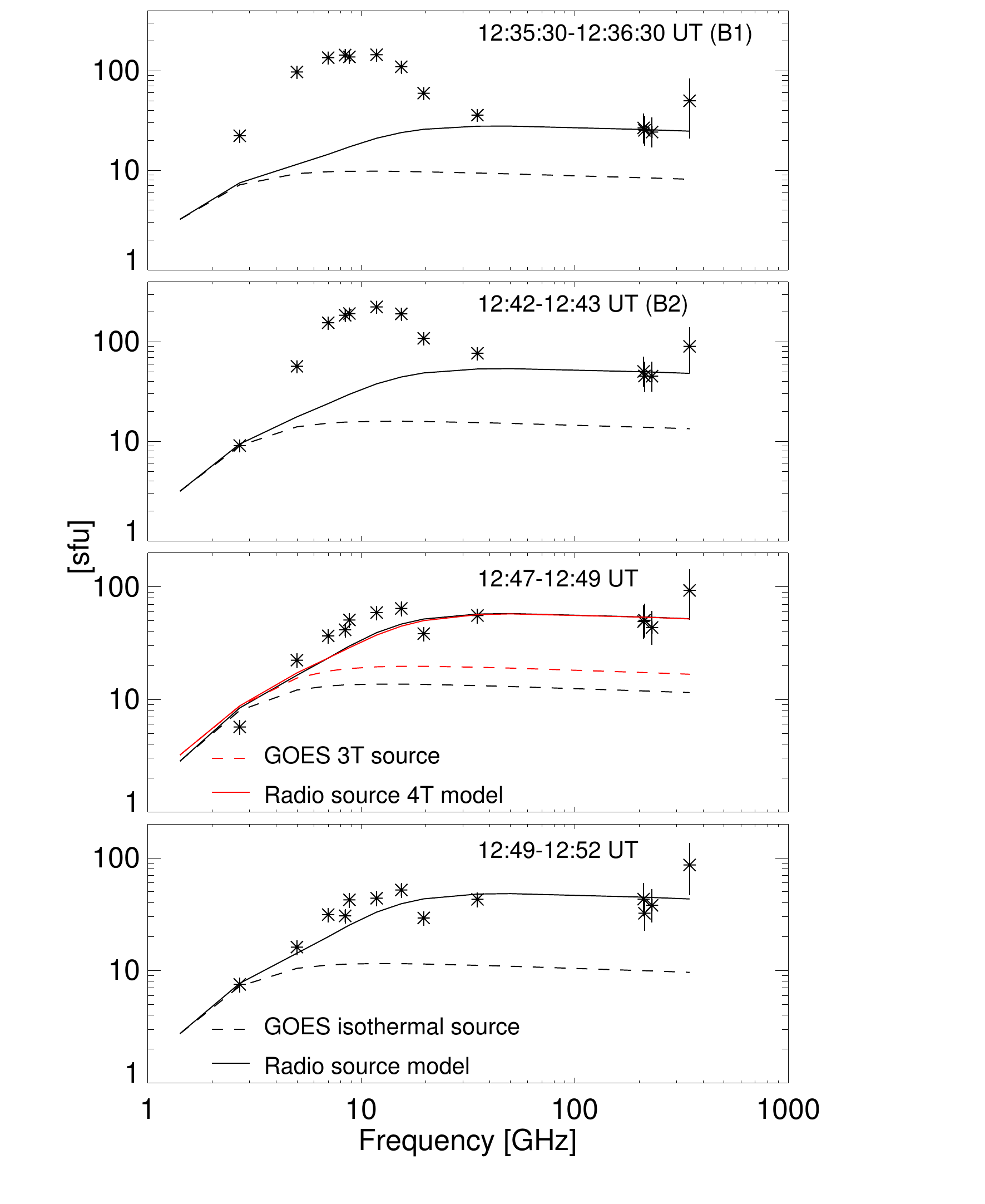}
}
\caption[]{
The mean radio spectrum observed at different times during the time-extended (gradual) phase of the 2003~October~27 flare (asterisks with errors bars). From top to bottom, the spectra correspond to bursts B1 and B2 (see Figure~\ref{fig_lcradio}), and to the plateau and decay of the 210--230~GHz emission, respectively. The dashed lines indicate the radio spectrum expected purely from an isothermal plasma with temperature and emission measure derived from the GOES SXR measurements, assuming a circular source of $40\arcsec$ diameter, while the solid lines indicate the spectrum expected from a two-temperature model (see text).}
\label{fig_spgrad}
\end{figure}

As stated in Section~\ref{Sec_Obs}, the slowly-varying and time-extended (``gradual'') radio emission starts to rise close to the onset of the SXR emission measured by GOES (see Figure~\ref{fig_lcradio}). Superimposed on this slowly-varying component, two impulsive bursts, labeled B1 and B2 on Figure~\ref{fig_lcradio}, are observed at frequencies up to 35~GHz but without obvious counterparts in the 200--400~GHz domain. Figure~\ref{fig_spgrad} shows the mean radio spectrum (asterisks) observed during B1 and B2 and during two later time intervals. For each of the four time intervals shown, the dashed curve indicates the mean bremsstrahlung spectrum expected from a single isothermal source of assumed $40\arcsec$ diameter with temperature $T_{\rm g}$ and emission measure $E\!M_{\rm g}$ as derived from the GOES SXR fluxes (see Section~\ref{Sb1_Gradual}). The main points to be drawn from Figure~\ref{fig_lcradio} and Figure~\ref{fig_spgrad} are:

\begin{itemize} 
	\item [--] During the two last intervals, when no impulsive bursts are present, the observed radio spectrum is almost flat between $\sim$8 and 230~GHz, with flux densities $\sim$3 to 4 times greater than those expected from the single thermal source derived from GOES SXRs.
	\\
	\item [--] During bursts B1 and B2, a gyrosynchrotron component is superimposed on a thermal bremsstrahlung spectrum. B1 has no counterpart in the mm--submm range, but the $>$200~GHz radiation exhibits an additional increase during B2.
	\\
	\item [--] Throughout the gradual phase, the 345~GHz flux density is significantly higher than that measured in the 210--230~GHz range (see Section~\ref{Sec_FluxKOSMA}).
\end{itemize}
We discuss each of these points in the following Sections.


\subsubsection{The Flat Radio Spectrum} 
 \label{Sb1_Gradual}

The flat spectrum observed between $\sim$8 and 230~GHz throughout the gradual phase is reminiscent of optically-thin thermal bremsstrahlung emission. Below $\sim$8~GHz, the flux density decreases with decreasing frequency, indicating that the radio source gradually becomes optically-thick at low frequencies. To characterize the thermal plasma seen by GOES, its  temperature $T_{\rm g}$ and emission measure $E\!M_{\rm g}$ were derived from the GOES SXR fluxes, assuming an isothermal source, using the CHIANTI database \cite{Der:al-97,You:al-98} and coronal elemental abundances. $T_{\rm g}$ peaks at $\sim$20~MK just after the maximum of the impulsive phase, then decreases to $\sim$12~MK. $E\!M_{\rm g}$ initially increases, passes through a maximum of $\sim$$3\times10^{49}$~cm$^{-3}$ approximately at the end of B2, and then falls to $\sim$10$^{49}$~cm$^{-3}$.

We computed the radio spectrum expected from the GOES thermal source as follows.  At time $t$ and frequency $\nu$, the brightness temperature $T_{\rm br}(t,\nu)$ of the radio emission from an isothermal, homogeneous source at temperature $T(t)$, emission measure $E\!M(t)$, and projected surface area $A$ (normal to the line-of-sight) is given by:
\begin{equation}
\label{Eq_Tb1}
	T_{\rm br}(t,\nu) = T(t) \left(1-e^{-\tau(t,\nu)}\right) \,,
\end{equation}
where, for $T>0.2$~MK, the optical thickness $\tau(t,\nu)$ is given by \cite{Dul-85}:
\begin{align}
\begin{split}
	\tau(t,\nu) &= \frac{K(t,\nu)}{\nu^2A}\frac{E\!M(t)}{T(t)^{3/2}} \,, \\
	K(t,\nu) &\equiv 9.78 \times 10^{-3} \times (24.5 +{\ln{T(t)}} - {\ln\nu})\quad {\rm (c.g.s.\ units)} \,.
\label{Eq_Tau}
\end{split}
\end{align}
The expected radio flux density $S_{\rm exp}(t,\nu)$ from the source is then given by:
\begin{equation}
\label{Eq_Flux}
	S_{\rm exp}(t,\nu) = \frac{2k_{\rm B}T_{\rm br}(t,\nu)\nu^2A}{c^2R^2} \,,
\end{equation}
where $c$ is the speed of light and $R$ is the Sun--Earth distance.  We note that for optically-thin ($\tau\ll1$) regimes, Equations~(\ref{Eq_Tb1}) and (\ref{Eq_Flux}) reduce to:
\begin{equation}
\label{Eq_Tbtn1}
	T_{\rm br}(t,\nu) \approx T(t)\tau(t,\nu) \; \implies \; S_{\rm exp}(t,\nu) = \frac{2k_{\rm B}K(t,\nu)}{c^2R^2}\frac{E\!M(t)}{T(t)^{1/2}} \,,
\end{equation} 
whereby the expected radio flux spectrum is independent of the source area $A$ and only weakly dependent on frequency $\nu$.

Using $T_{\rm g}(t)$ and $E\!M_{\rm g}(t)$ derived for the GOES thermal plasma, and assuming a projected area equivalent to a circle of $40\arcsec$ diameter (the mean acceptable source area, per Section \ref{Sec_FluxSST}; see also the discussion below), we calculated the radio flux density $S_{\rm exp}(t,\nu)$ expected for this source, shown by the dashed black curves in Figure~\ref{fig_spgrad}. Like \inlinecite{Lut:al-04}, we find a qualitative similarity between the time evolutions of the observed and predicted flux densities in the 200--400~GHz range (where the emission is optically thin and where no impulsive bursts are detected), but the observed values are $\sim$3--4~times higher at 210--230~GHz and $\sim$7~times higher at 345~GHz. Even using photospheric rather than coronal abundances to derive $T_{\rm g}$ and $E\!M_{\rm g}$, the computed radio fluxes are still $\sim$2 and $\sim$4 times lower than the observed values at 8--230 and 345~GHz, respectively.

In contrast, \inlinecite{Poj:al-96} reported gradual radio bursts at 37~GHz for which the observed and computed flux excesses were comparable. A similar agreement was obtained from case studies at 8.8 and 15.4~GHz by, e.g., \inlinecite{Kun:al-94}, up to 15.4~GHz by \inlinecite{Tro:al-02}, and up to 86~GHz by \inlinecite{Rau:al-99}. However, we note that in these earlier studies, $T_{\rm g}$ and $E\!M_{\rm g}$ were derived using the \inlinecite{Tho:al-85} polynomial approximation, which closely resembles the GOES response to {\it photospheric}\hyp{}abundance CHIANTI models for temperatures up to $\sim$15~MK \cite{Whi:al-05}. The GOES plasma likely resides in coronal loops, not chromospheric footpoints, and as shown above, the use of photospheric abundances results in substantially larger expected radio flux densities than those obtained using the more appropriate coronal abundances.  Thus, this suggests that, for most of the gradual bursts reported in the cited works, the measured flux densities in the centimeter--millimeter range are actually higher than those expected from the SXR-emitting plasma observed by GOES, similar to our results.

\inlinecite{Whi:Kun-92} emphasized that the above discrepancy between observed radio emission and that computed from SXR observations mainly reflects that: (i)~the assumption of an isothermal coronal plasma is not correct during a flare, and (ii)~radio observations are more sensitive to cooler material than is GOES because, in contrast to SXR intensity, the intensity of optically-thin radio emission decreases with increasing temperature. They concluded that a 0.1--1~MK source would produce a stronger response in the centimeter--millimeter domain than would (hotter) plasma observed by GOES. \inlinecite{Che:al-95} reached similar conclusions in a joint analysis of GOES observations and radio bursts measured at 3--80~GHz.

Following these authors, we consider that, in addition to the coronal SXR source SG observed by GOES, there is an additional, cooler, radio-emitting coronal source SR at plasma temperature $T_{\rm r} < T_{\rm g}$ with emission measure $E\!M_{\rm r}$. For simplicity, we assume that SR has the same surface area $A$, normal to the line of sight, as SG and that it is located directly below it. Then, by modifying Equation~(\ref{Eq_Tb1}) for this geometry, we obtain the brightness temperature of the combined source: 
\begin{equation}
\label{Eq_Tb}
{T_{\rm br}(t,\nu) =	T_{\rm g}(t) \left(1-e^{-\tau_{\rm g}(t,\nu)}\right) +
				T_{\rm r} (t) \left(e^{-\tau_{\rm g}(t,\nu)} - e^{-\left(\tau_{\rm g}(t,\nu) + \tau_{\rm r}(t,\nu)\right)}\right)} \,,
\end{equation}
which, together with Equations (\ref{Eq_Tau}) and (\ref{Eq_Flux}), yield the expected coronal flux density $S_{\rm exp}^{\rm cor}(t,\nu)$ for the combined source.  If both SG and SR are optically thin ($\tau_{\rm g} \ll 1$ and $\tau_{\rm r} \ll 1$), Equations~(\ref{Eq_Tb}) and (\ref{Eq_Flux}) reduce to:
\begin{align}
\begin{split}
	T_{\rm br}(t,\nu) &\approx T_{\rm g}(t)\tau_{\rm g}(t,\nu) + T_{\rm r}(t)\tau_{\rm r}(t,\nu) \,, \\
	\implies \; S_{\rm exp}^{\rm cor}(t,\nu) &= \frac{2k_{\rm B}}{c^2R^2}\left(\frac{K_{\rm g}(t,\nu)E\!M_{\rm g}(t)}{T_{\rm g}(t)^{1/2}}+\frac{K_{\rm r}(t,\nu)E\!M_{\rm r}(t)}{T_{\rm r}(t)^{1/2}}\right) \,.
\label{Eq_Tbtn}
\end{split} 
\end {align}
$T_{\rm g}$ and $E\!M_{\rm g}$ are fixed by the GOES observations; the free parameters $T_{\rm r}$, $E\!M_{\rm r}$, and $A$  can then be determined by fitting the model $S_{\rm exp}^{\rm cor}(t,\nu)$ (Equation~(\ref{Eq_Flux})) to the whole range of  observed flux densities $S_{\rm obs}(t,\nu)$.

As a first approach, we varied the free parameters of the cool plasma SR until a reasonable fit to the data was achieved by eye. We constrained the fit parameters by requiring that the cool plasma does not contribute significantly to the GOES SXR flux in either energy channel, since that would conflict with the previously-derived $T_{\rm g}$ and $E\!M_{\rm g}$ which were held fixed in the model.  In this way, we determined the best-fit parameters of the cool plasma for 1-minute time intervals from 12:33 to 12:57~UT; during B1 and B2, when synchrotron radiation adds to the thermal bremsstrahlung up to at least 35~GHz, $T_{\rm r}$ and $E\!M_{\rm r}$ were adjusted to account for the observed 210--230~GHz flux densities. 

Figure~\ref{fig_spgrad} illustrates the reasonable agreement of the two-temperature coronal source model (solid black line) with the observations (asterisks) for $A = 6.7 \times 10^{18}$~cm$^{2}$, which corresponds to a diameter of $40\arcsec$ for a uniform circular source.  $T_{\rm r}$ and $E\!M_{\rm r}$, respectively, begin at 1~MK and $7 \times 10^{48}$~cm$^{-3}$ in the first time interval (12:33--12:34~UT), reach a maximum of 2.6~MK and $3.5 \times 10^{49}$~cm$^{-3}$ in the 12:43--12:45~UT time period (decay of burst B2), and decrease to 1.3~MK and $1.2 \times 10^{49}$~cm$^{-3}$ at the end of the analyzed period (12:57~UT). This $40\arcsec$, 1--3~MK plasma, with $E\!M_{\rm r} \approx E\!M_{\rm g}$, contributes $<$1\% of the GOES SXR flux in either channel, as we required.  Larger sources, up to $60\arcsec$, yield similar values of $T_{\rm r}$ and $E\!M_{\rm r}$, but smaller sources yield higher temperatures and emission measures, resulting in a non-negligible predicted contribution of the SR source to the GOES flux in at least the low-energy channel. A $30\arcsec$ source, for example, with best-fit $T_{\rm r} \approx 4.6$~MK and $E\!M_{\rm r} \approx 5 \times 10^{49}$~cm$^{-3} \approx 2 \times E\!M_{\rm g}$, would contribute only $\sim$0.5\% of the SXR flux at 0.5--4~{\AA} but a significant $\sim$16\% at 1--8~{\AA}, which violates our initial constraints. Thus, we adopt a coronal radio source diameter $D_{\rm r} =  40\arcsec \pm 10\arcsec$.

\begin{figure}
\centerline {
	\includegraphics[width=0.9 \textwidth]{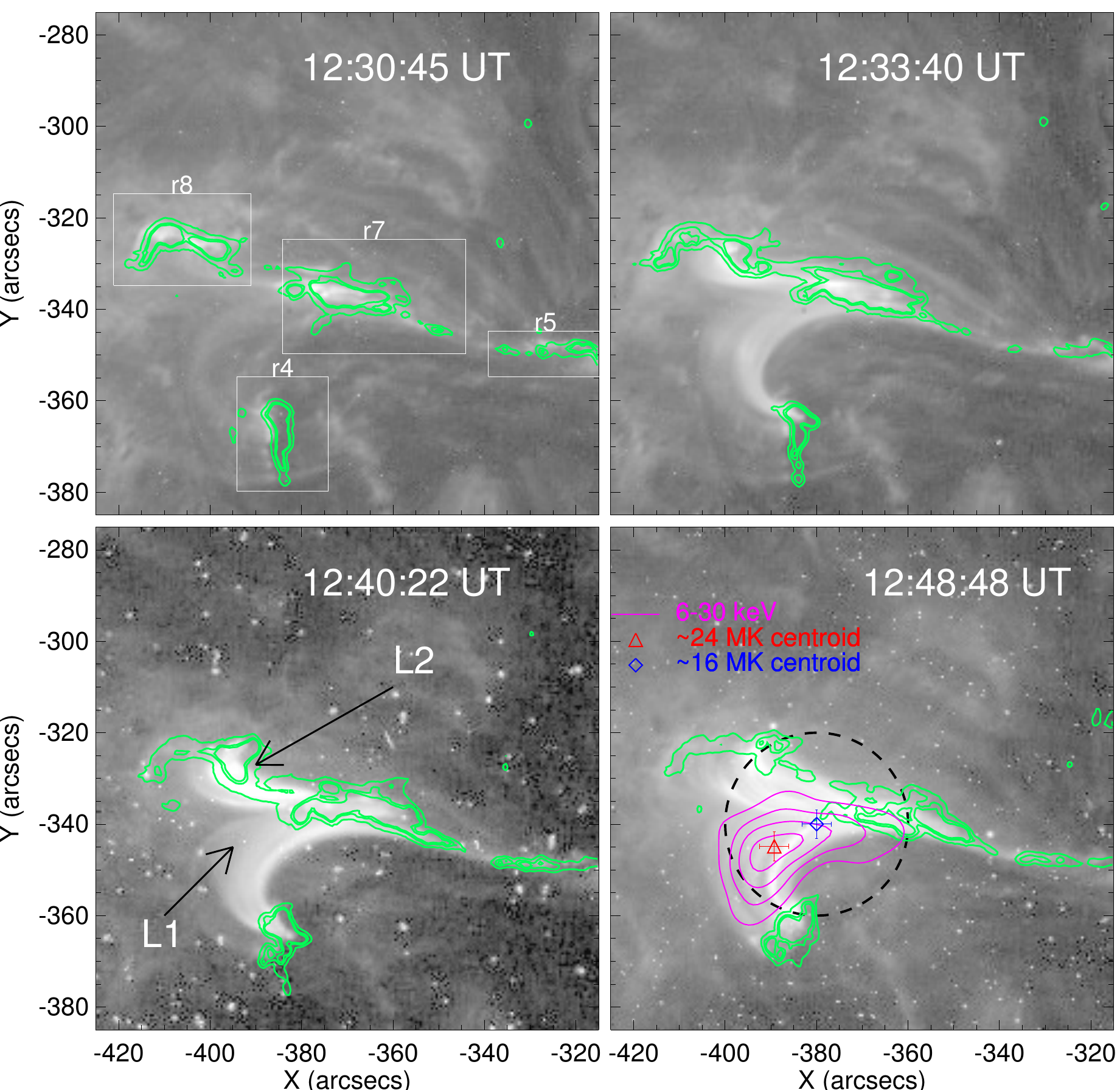}
}
\caption[]{TRACE images at 195~{\AA} (reverse grayscale) and 1600~{\AA} (green contours) at four instants during the 2003~October~27 flare; the TRACE alignment was corrected by [+6\arcsec, -7.5\arcsec]. The 12:30:45~UT image (top left) occurs during the impulsive phase; r4, r5, r7 and r8 refer to the 1600~{\AA} flare ribbons/kernels as identified in Figure~4 of \inlinecite{Luo:al-07}, and the gray boxes indicate the areas over which the kernel intensities have been computed (see Sec.~\ref{Sb3_Gradual}).  The three subsequent images, which exhibit two loop systems marked L1 (south) and L2 (north), occur during the gradual phase.  RHESSI 6--30 keV image contours (30\%, 50\%, 70\%, 90\%) during 12:47--12:49~UT are overlaid on the final image (bottom right), along with the derived centroid positions of the two thermal plasmas inferred from the simultaneous, spatially-integrated RHESSI spectrum (see Sec.~\ref{Sec_RHESSI}). For comparison, the dashed black circle of 40\arcsec\ diameter represents the model coronal radio source, arbitrarily centered on the centroid of the $\sim$16~MK plasma.}
\label{fig_trim}
\end{figure}

While this simple two-temperature model prediction of the radio spectrum yields reasonable agreement with the observations, the RHESSI data, when available, provide distinct evidence that the true temperature distribution of the thermal flare plasma is more complex.  For the specific time interval of 12:47--12:49~UT (see Section~\ref{Sec_RHESSI}), the RHESSI $\sim$3--33 keV spectrum is well characterized by two hot, isothermal components with $T_{1}\approx24$~MK, $E\!M_{1}\approx0.24\times10^{49}$~cm$^{-3}$, $T_{2}\approx16$~MK, and $E\!M_{2}\approx1.7\times10^{49}$~cm$^{-3}$.  Because of the broadband GOES response, the $\sim$24~MK source will contribute non-negligibly to the measured SXR flux, which will therefore affect the radio spectrum predicted from the GOES measurements.  In particular, the two thermal components observed by RHESSI contribute a combined 85\% and 98\% of the GOES SXR flux observed in the low- and high-energy channels, respectively. The remaining GOES fluxes are consistent with an additional ``warm'' plasma, with $T_{3} \approx 6.7$~MK and $E\!M_{3} \approx 1.0 \times 10^{49}$~cm$^{-3}$.  However, the red dashed curve in Figure~\ref{fig_spgrad} shows that the radio emission predicted from this three-temperature ($T_{1}$, $T_{2}$, $T_{3}$) plasma is still below the observed spectrum.  Thus, a fourth, cool plasma, with $T_{4} \approx 2.0$~MK and $E\!M_{4} \approx 3.0 \times 10^{49}$~cm$^{-3}$, is still needed to account for the observed radio flux densities (with a diameter $D_{\rm r}=40\arcsec\pm10\arcsec$, as before); this cool plasma contributes negligibly ($<$1\%) to the GOES fluxes, as required.

Figure~\ref{fig_spgrad} shows that the four-temperature model prediction (solid red line in the 12:47--12:49~UT panel) is nearly identical to that of our previous, simple, two-temperature model (solid black line). Thus, in  the absence of RHESSI data, a simple two-temperature scenario nevertheless constitutes a viable working model, representative of the radio spectrum from a multi-temperature coronal plasma.  However, we note that the derived $T$ and $E\!M$, and hence the predicted radio contribution, of the cool source do vary between the two models, so additional measurements within the temperature distribution are desirable for optimal accuracy, when possible.

\begin{sloppypar}TRACE observations at 195 and 1600~{\AA} were made between 12:28 and 13:00~UT, with a few observations at 284~{\AA} also made after 12:41~UT. Images at 195~{\AA} (and at 284~{\AA}) show the geometry of the flaring coronal plasma, while the 1600~{\AA} images display the chromospheric footpoints/ribbons, as illustrated in Figure~\ref{fig_trim} at four instants during the event. The bright 1600~{\AA} kernels are marked r4, r5, r7, and r8, as in Figure~4 of \inlinecite{Luo:al-07}; similar kernels are also seen in H$\alpha$ (see Figure~3, right panel, of \opencite{Luo:al-07}). The image at 12:30:45~UT, characteristic of the impulsive phase, shows only faint 195~{\AA} emission directly atop the 1600~{\AA} kernels. During the gradual phase, Figure~\ref{fig_trim} shows two bright 195~{\AA} loop systems, marked L1 and L2, which connect the 1600~{\AA} kernels. The 195~{\AA} intensity, integrated over the area covered by these loops, increases from the beginning of the gradual phase up to at least 12:40~UT; after that time, the images are contaminated by numerous spikes that prevented a reliable measure of the light curve.\end{sloppypar}

The 195~{\AA} passband is sensitive to spectral line emission from both Fe~{\sc xii} (peak contribution at $\sim$1.1--1.7~MK, cf. \opencite{Maz:al-98})  and Fe~{\sc xxiv} ($\sim$13--35~MK), and the 195~{\AA} images will thus contain some contribution from the hot $\sim$24 and $\sim$16~MK plasmas seen by RHESSI.  However, per unit emission measure, the sensitivity is $\gsim$10 times stronger at the low (Fe~{\sc xii}) temperatures (cf. \opencite{Han:al-99}), and thus the total 195~{\AA} emission in L1 and L2 should be vastly dominated by cool plasma.  This is confirmed by the similar loop structures observed at 284~{\AA} (Fe~{\sc xv}, $\sim$1.8--2.6~MK), and suggests that L1 and L2 are indeed largely representative of the derived 1--3~MK source considered in our working model, and therefore of the coronal thermal radio-emitting region. The increase of the 195~{\AA} intensity up to at least 12:40~UT is thus consistent with our model, which predicts an increase of $T_{\rm r}$ and $E\!M_{\rm r}$ until 12:43--12:45~UT.

The spatial distribution of the RHESSI-observed 6--30~keV emission during 12:47--12:49~UT, where we applied the four-temperature thermal model, is also shown in Figure~\ref{fig_trim} (bottom right panel). The X-ray source has a similar shape as and projects onto the L1 loop system. The projected centroid positions of the $\sim$24 and $\sim$16~MK plasmas (see Section~\ref{Sec_RHESSI}) are located near the top of the brightest loops within L1, as would be expected for the hottest plasma (cf. \opencite{Cas:Lin-10}). The projected (and deconvolved) surface area of the RHESSI X-ray source, $\sim$300 and $\sim$600 square arcsec within the 50\% and 30\% contours, respectively, is significantly smaller than the $\sim$1250 square arcsec area of the radio source obtained from the model.  Since the X-ray source produces only $\lsim$20\% of the radio flux (see Figure~\ref{fig_spgrad}), its size is not likely representative of the primary radio source, suggesting that the 4-$T$ model assumption of identical sizes for the hot ($>$16~MK), warm ($\sim$6--7~MK), and cool ($\sim$1--3~MK) sources may not be valid. However, this does not affect the results in the optically-thin regime, which are independent of source size (cf. Equation (\ref{Eq_Tbtn})). Although the 195~{\AA}-emitting region (bounded by a $\sim$$60\arcsec \times 50\arcsec$ rectangle) is very inhomogenous, with a diffuse edge that is difficult to define precisely, we estimate that the projected surface area is between $\sim$800 and $\sim$1150 square arcsec.  While the areas of the X-ray and EUV sources are estimated using dissimilar methods, the difference in their areas is significant, and it is noteworthy that the X-ray source has negligible emission overlying the L2 system, which can largely account for its smaller size.  The 195~{\AA} source area is, on the other hand, in reasonable agreement with the $\sim$1250 square arcsec area of the model radio source, consistent with the model finding that the radio emission is primarily from the $\sim$1--3~MK plasma observed by TRACE.  Unfortunately, the present data do not provide enough constraints to consider a more sophisticated model which accounts for inhomogeneities within the EUV- and X-ray-emitting regions; observations from UV/EUV and X-ray imagers and spectrometers onboard the {\it Solar Dynamics Observatory} and {\it Hinode} satellites may provide the necessary information to develop such complex and realistic models in the future.

In summary, the 1--230~GHz radio spectrum observed during the gradual phase is consistent with thermal bremsstrahlung radiated by a coronal source of $\sim$$40\arcsec$ equivalent diameter, filled with hot, $\sim$7--24~MK plasma observed by GOES and RHESSI and with cooler, $\sim$1--3~MK plasma observed by TRACE. The similarity between the time evolution of the SXR and radio emissions, particularly in the 200--400~GHz range (see Figure~\ref{fig_lcradio}), implies that these emissions likely share a common origin. The increase of $E\!M_{\rm g}$ and $E\!M_{\rm r}$ from the beginning to the maximum of the gradual phase suggests that this common origin could be chromospheric evaporation.

Since the radio corona is optically-thin above $\sim$8~GHz, the observed flux density $S_{\rm obs}$($\nu$) at frequency $\nu > 8$~GHz can be written as the sum of the emission from the corona, $S_{\rm obs}^{\rm cor}(\nu)$, and from the chromosphere, $S_{\rm obs}^{\rm chr}(\nu)$. Similarly, the flux density $S_0$ before the flare is thus $S_0^{\rm cor} + S_0^{\rm chr}$.
Then, the excess flux density due to the flare, $S_{\rm F}$, is given by:
\begin{equation}
\label{Eq_Exfla}
{S_{\rm F}(\nu) = S_{\rm obs}^{\rm cor}(\nu) - S_0^{\rm cor}(\nu) + S_{\rm obs}^{\rm chr}(\nu) - S_0^{\rm chr}(\nu)} \,.
\end{equation}
The observation of optically-thin bremstrahlung during the gradual phase leads to the following statements: (i)~the excess flux seen at 345~GHz and the enhanced 1600~{\AA} and H$\alpha$ emissions (see Section~\ref{Sb3_Gradual}, below) provide evidence that the chromosphere responded markedly to flare energy deposition, but (ii)~Equation~(\ref{Eq_Exfla}) implies that, since the observed flux densities are well-modeled entirely by coronal sources (per above), $S_{\rm obs}^{\rm chr}(\nu) \approx S_0^{\rm chr}(\nu)$; that is, the chromosphere did not contribute substantially to the time-extended radio emission over a large frequency range $\Delta\nu \approx 10$--230~GHz. This range may vary from flare to flare, and even during a single flare, such as was the case during the 2001~April~12 event \cite{Lut:al-04} for which $\Delta\nu$ varies from $\sim$10--230~GHz at the maximum of the thermal phase to $\sim$10--100~GHz in the late decay.  Variations of $\Delta\nu$ during the gradual phase likely reflect the dynamics of the chromospheric response to flare energy deposition.


\subsubsection{The Bursts B1 and B2}
\label{Sb2_Gradual}

Although a detailed study of bursts B1 and B2 is not of interest here, it is necessary to identify the origin and the nature of these bursts to understand the relationship between the radio,  UV and H$\alpha$ emission discussed in Section~\ref{Sb3_Gradual}, below.
 
 B1 and the smaller burst just prior to it (see Figure~\ref{fig_lcradio}) are both observed between 5 and 35~GHz in time-coincidence with Type~III electron beams recorded by NRH. NRH imaging in the metric domain shows that these Type~IIIs arise from AR~10484 at N07~W46. Furthermore, high cadence H$\alpha$ images, obtained with the {\it H$\alpha$ Solar Telescope for Argentina} (HASTA; \opencite{Bag:al-99}; \opencite{Fernandez-02}), also show simultaneous brightenings in the same active region. We thus conclude that B1 is not part of radio emission arising from the 2003~October~27 flare under study.

In contrast, B2, which is also observed in the 5--35~GHz range, does arise from AR~10486. Indeed, Figure~\ref{fig_lckern} shows that B2 is coincident with an impulsive increase of the four 1600~{\AA} kernels marked r4, r5, r7 and r8 in Figure~\ref{fig_trim}. The time-derivative of the GOES 1--8~{\AA} SXR flux, which may be considered as a proxy for non-thermal hard X-ray emission (cf. the ``Neupert effect;'' \opencite{Neu-68}; \opencite{Den:Zar-93}), also exhibits a bump coincident with B2 (see Figure~\ref{fig_lckern}). As both non-thermal X-rays and the 1600~{\AA} UV continuum are widely believed to result from direct particle injection into the chromosphere \cite{Fle:Hud-01,Coy:Ale-09}, B2 is thus the signature of a new episode of particle acceleration in the AR~10486 magnetic field. After subtraction of the thermal radio spectrum expected from the model described in Section~\ref{Sb1_Gradual}, the remaining B2 spectrum is a typical gyrosynchrotron spectrum with a maximum at $\sim$10~GHz and a power-law spectral index of approximately $-2$ above that frequency. For such an index, the expected flux density above 200~GHz is $<$1~sfu at the maximum of B2. This is consistent with the lack of detection of synchrotron emission from B2 at these high frequencies. However, the $>$200~GHz emission shows a significant increase from the beginning of B2 until slightly after its maximum. This is an indication that the new energy release associated with B2 is not only a signature of a new episode of particle acceleration, but that it also produced a significant increase of the thermal radio emission measured during the gradual phase.


\subsubsection{The flux density increase between 210-230 and 345 GHz}
\label{Sb3_Gradual}

\begin{figure}
\centerline {
	\includegraphics[width=0.9 \textwidth]{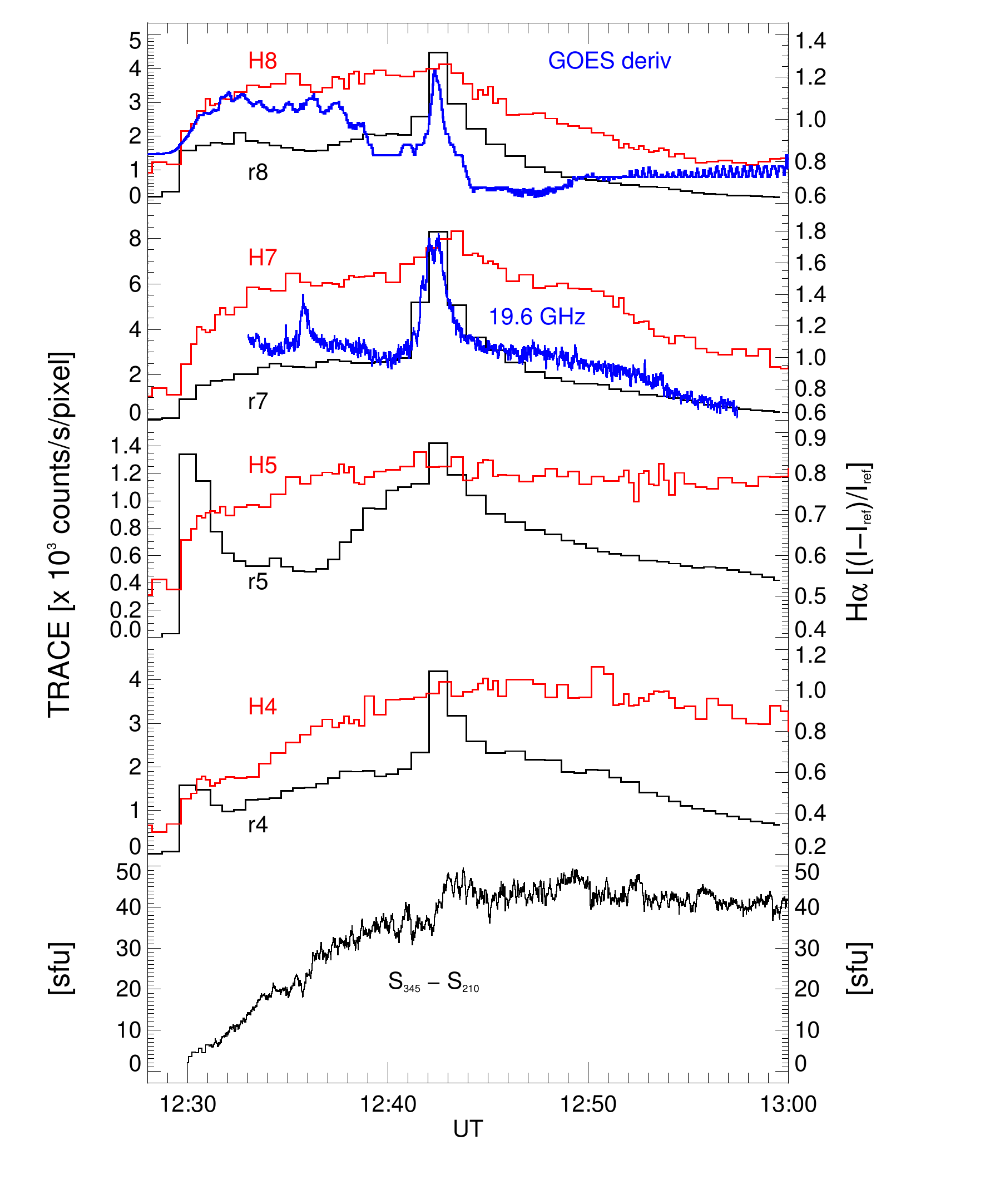}
}
\caption[]{From bottom to top: time evolution of the excess flux density at 345~GHz with respect to that of the flux density at 210~GHz; and time evolution of the four 1600~{\AA} kernels marked r4, r5, r7 and r8 on Figure~\ref{fig_trim} (black curves) and of the corresponding H$\alpha$ kernels H4, H5, H7 and H8 (red curves). At 1600~{\AA}, the pre-flare intensity of the kernel has been subtracted. The intensity excess of a given H$\alpha$ kernel $I-I_{\rm ref}$ is given with respect to that of a quiet area $I_{\rm ref}$. The blue curves overplotted on the time histories of kernels r7 (H7) and r8 (H8) are, respectively, the time evolutions of the 19.6~GHz radio emission and of the time-derivative of the GOES 1--8~{\AA} SXR flux. }
\label{fig_lckern}
\end{figure}

\inlinecite{Luo:al-07} noted that the 195~{\AA}-emitting loops shown in Figure~\ref{fig_trim}, with footpoint regions r4, r5, r7, and r8, extended above the inversion lines separating an elongated region of negative polarity from positive polarities northward and southward. Similar kernels, which we label H4, H5, H7, and H8, are also seen in the H$\alpha$ line center. \citeauthor{Luo:al-07} suggested that the flare energy release resulted from magnetic reconnection at a coronal null located at a height of 3.1~Mm, slightly to the east of r7 (see their Figure 6). In order to investigate the chromospheric response to this energy release, we analyzed the time evolution of the 1600~{\AA} and H$\alpha$ kernels. For this, we computed the mean intensities per pixel over the boxes shown in Figure~\ref{fig_trim}. Figure~\ref{fig_lckern} shows the results for both the 1600~{\AA} kernels (black histograms) and the H$\alpha$ kernels (red histograms). The bottom panel shows the excess flux at 345~GHz with respect to the 210~GHz flux ($\delta S_{345} \equiv S_{345} - S_{210}$). The examination of Figure~\ref{fig_lckern} suggests the following:
\begin{itemize}
	\item [--] Kernels r4 and r5 show similar time behaviors. Indeed, they both exhibit a short peak corresponding to the radio impulsive phase, another peak corresponding to the radio burst B2, and similar time-decays. H4 and H5 also evolve similarly, but unlike for r4 and r5, the impulsive phase and B2 are not seen in their time profiles. We emphasize that the H4 and H5 intensities remain roughly constant from $\sim$12:42~UT until the end of the analyzed period.
	\\
	\item [--] There is a close similarity between the r7 and r8 light curves. The marked difference compared to r4 and r5 is that the impulsive phase is not clearly observed, and their decay times are slightly shorter. H7 and H8 evolve together, but unlike H4 and H5, they decay after $\sim$12:43~UT. It should be noted that the region in which kernel r7 (H7) is located is highly dynamic. The elongated shape of this kernel is indicative of the photospheric trace of the fan associated with the null point found in \inlinecite{Luo:al-07}. Magnetic reconnection at the null point would inject energy into the chromosphere at the footpoints of loops anchored in r7, but at the same time, would induce a disturbance at the section of a long, curved filament lying along the inversion line located in its close vicinity (see, e.g., Figure 2 in \opencite{Mandrini-06}). The temporal evolution of the pixels included in the box shown in Figure~\ref{fig_trim} may thus reflect not only the chromospheric response at the loop footpoints but also the nearby filament activity. Nevertheless, we computed the r7 intensity for different sub-boxes within the box shown in Figure~\ref{fig_trim} and found that the shapes of the time profiles are not substantially different from the one shown in Figure~\ref{fig_lckern}.
\end{itemize}

The above observational findings therefore support the following statements:

\begin{itemize}
	\item [--] After reconnection and energy release at the magnetic null point, the two sets of 195~{\AA} loops L1 and L2 have footpoints at r4 (H4) and r5 (H5) and at r7 (H7) and r8 (H8), respectively (see Figure~\ref{fig_trim}). Taking into account the magnetic field evolution (e.g., emergence of a new bipole into the pre-existing field configuration), \inlinecite{Luo:al-07} proposed that magnetic reconnection at the null point would proceed in such a way that field lines connecting r4 to r7 and r5 to r8 would correspond to L1 and L2, contrary to what our analysis of the temporal evolution of the flare kernels show. Note that it is very difficult to clearly separate these two loops in the horizontal ``Y-shape'' of the TRACE 195~{\AA} brightening. However, computation of the magnetic field topology is only an indication of the magnetic field connectivity and does not tell in which direction the reconnection process proceeds; this is inferred by analyzing the temporal evolution of the associated phenomena. Furthermore, magnetic reconnection may proceed in one direction and, afterwards, in the reverse direction in consecutive events, as was shown by \inlinecite{Goff-07}. Therefore, we conclude that the magnetic field topology computed by \inlinecite{Luo:al-07} is in agreement with the connectivities inferred from our analysis of TRACE and H$\alpha$ data, although the reconnection process may have proceeded in the reverse sense compared to that proposed by these authors. 
	\\
	\item [--] As stated in Section~\ref{Sb2_Gradual}, it is well-documented that the 1600~{\AA} peaks, associated with the impulsive phase and B2, constitute signatures of the interaction of flare-accelerated particles with the chromospheric footpoints of loops connected to the acceleration/injection region. Our results thus indicate that, during the impulsive phase, particles are preferentially injected into L1, while during B2, they are injected into both L1 and L2.
	\\
	\item [--] The time profile of $\delta S_{345}$ closely mimics the H4 and H5 light curves. This strongly suggests that $\delta S_{345}$ is optically-thick thermal bremsstrahlung from the footpoints of L1. Assuming that an equal amount of 345~GHz emission arises from two sources of diameter $d$ (expressed in arcsec), the projections of which correspond to H4 and H5, we find that $\delta S_{345}$ leads to a temperature increase of up to $8 \times 10^3 \times (20\arcsec/d)^2$~K. For $d=10\arcsec$ to $20\arcsec$, i.e., comparable to H4 and H5 (see Figure~\ref{fig_trim}), this yields maximum temperature increases of $3.2 \times 10^4$ and $8 \times 10^3$~K, respectively. These temperatures are consistent with a chromospheric origin for $\delta S_{345}$ and with its similarity to the H$\alpha$ time profiles, the latter being the signature, when seen in emission as during our event, of $\sim$10$^4$~K plasma. 
\end{itemize}

The latter point above and the discussion in Section~\ref{Sb1_Gradual} suggest that, during the gradual phase of the flare, the 230~GHz emission arises from a 1--3~MK source of $40\arcsec$ diameter centered near the top of L1, while the 345~GHz emission arises from both the same coronal source and from two chromospheric sources of $\sim$$20\arcsec$ diameter that project onto r4 (H4) and r5 (H5). The convolution of this configuration of sources with the KOSMA beams (HPBWs of $117\arcsec \pm 7\arcsec$ at 230~GHz and $88\arcsec \pm 5\arcsec$ at 345~GHz, cf. \opencite{Lut:al-04}) yields source sizes of $124\arcsec$ at 230~GHz and $107\arcsec \times 96\arcsec$ at 345~GHz. If we then assume that the emission comes from unresolved single sources, after deconvolution we find source sizes of $40\arcsec$ at 230~GHz, as expected, and $37\arcsec \times 61\arcsec$ at 345~GHz, substantially larger than the coronal 230~GHz source. Such an increase of source size with frequency was previously reported by \inlinecite{Lut:al-04}, who, during the late decay of the gradual phase of the 2001~April~12 flare, measured sizes of $42\arcsec \pm 20\arcsec$ and $70\arcsec \pm 6\arcsec$ at 230 and 345~GHz, respectively. A similar explanation as the one proposed here may qualitatively apply to the 2001~April~12 event. Indeed, Figure 4 of \inlinecite{Lut:al-04} indicates that the 345~GHz-emitting source encompasses both of the two chromospheric footpoints seen in hard X-rays, while the 230~GHz source encompasses only the northern one.


\section{Conclusions}

The 2003~October~27 M6.7 flare at $\sim$12:30~UT is one of the few events for which a time-extended phase (tens of minutes) following a short (few minutes) impulsive phase has been well observed at mm--submm wavelengths. The radio data from a few tens of kHz up to 405~GHz shows that the event was confined in the chromosphere and low corona, confirming earlier conclusions from the study of the active region magnetic field topology. Consistent with the confined nature of this event,  the 1--345~GHz emission detected during the impulsive phase is produced by gyrosynchrotron radiation from high-energy electrons in a dense medium.  

The combined analysis of radio total-flux observations, X-ray imaging and spectral data, along with UV/EUV and H$\alpha$ spatially-resolved  observations, has allowed us to determine the nature and origin of the radio emission during the gradual phase of the flare. The main findings can be summarized as follows:

\begin{itemize}
	\item [--] Except for the small impulsive burst B2 where gyrosynchrotron emission contributes up to 35~GHz, the radio emission is entirely produced by thermal bremsstrahlung.
	\\
	\item [--] Below 230~GHz, the radio emission, which is optically-thin above $\sim$8~GHz, is entirely produced in the corona by hot and cool materials at $\sim$7--16~MK and at $\sim$1--3~MK, respectively. The combination of X-ray data and EUV images at 195 and 284~{\AA} indicates that the $\sim$1--3~MK plasma fills two large coronal loop structures, L1 and L2, while the hot X-ray source lies primarily within L1, with the centroid of the $\sim$16~MK plasma near the top of the brightest portion of the L1 loop system.
	\\
	\item [--] At 345~GHz, in addition to the coronal emission, there is an optically-thick component arising from the lower atmosphere. The similarity between the time-evolution of this submm excess and that of the H$\alpha$ footpoints of L1 suggests that the excess 345~GHz flux density excess is emitted by chromospheric material at a few 10$^4$~K located in the L1 footpoints.
\end{itemize}

Importantly, these results show that  the chromospheric response to the energy release during the gradual phase produces negligible emission over a substantially\hyp{}extended frequency range, from $\sim$10 to $\sim$230~GHz for the studied flare; similar behavior has been observed in other flares, as well.  The thermal radio emission expected from semi-empirical flare models (cf. \opencite{Mac:al-80}; \opencite{Mau:al-90}) is inconsistent with the non-detection of chromospheric radio signatures over such a wide frequency range. Indeed, since the density--temperature structure considered in these models varies quasi\hyp{}monotonically with height, one would expect the radio flux density to increase with frequency in the spectral range where the corona is transparent.  Thus, observations of the radio continuum at cm--submm wavelengths and in the far infrared domain can serve as powerful diagnostic tools to investigate both the structure and the dynamics of the low solar atmosphere during flares.

\begin{acks}
GT acknowledges the FAPESP agency for support during his stay in Brazil (Proc. 2009/15880-0). 
JPR thanks CNPq (Proc. 305655/2010-8) and GGC is grateful to FAPESP (Proc. 2009/18386-7).
CHM and MLL acknowledge financial support from the Argentinean grants UBACyT X127, PIP 2009-100766 (CONICET), and PICT 2007-1790 (ANPCyT). CHM is a member of the Carrera del Investigador Cient\'\i fico (CONICET). MLL is a member of the Carrera del Personal de Apoyo (CONICET).
AC was supported by NASA grant NNX08AJ18G and NASA contract NAS5-98033.  
Thanks to E. Correia for providing 7~GHz polarimeter data. 
The authors wish to thank G. Hurford and K.-L. Klein for helpful discussions and comments.
We are grateful to the referee, S\"am Krucker, for his constructive recommendations. \end{acks}

\bibliographystyle{spr-mp-sola}

\bibliography{bib27oct03_new} 

\IfFileExists{\jobname.bbl}{} {\typeout{}
\typeout{****************************************************}
\typeout{****************************************************}
\typeout{** Please run "bibtex \jobname" to obtain}
\typeout{**the bibliography and then re-run LaTeX}
\typeout{** twice to fix the references !}
\typeout{****************************************************}
\typeout{****************************************************}
\typeout{}}

\end{article} 

\end{document}